%% file: 00-main.tex
\documentclass[conference]{IEEEtran}
\IEEEoverridecommandlockouts

\usepackage{balance}
\usepackage{cite}
\usepackage{amsmath,amssymb,amsfonts}
\usepackage{algorithmic}
\usepackage{graphicx}
\usepackage{textcomp}
\usepackage[dvipsnames]{xcolor}
\usepackage{soul}
\usepackage{booktabs}
\usepackage{multirow}
\usepackage{multicol}
\usepackage{subcaption}
\usepackage{algorithm2e}
\def\BibTeX{{\rm B\kern-.05em{\sc i\kern-.025em b}\kern-.08em
    T\kern-.1667em\lower.7ex\hbox{E}\kern-.125emX}}

\usepackage[detect-all=true, binary-units=true]{siunitx}

\usepackage[nolist,nohyperlinks]{acronym}
\usepackage{adjustbox}
\input{acronym}

\newcommand{\MYheader}{2022 International Conference on Localization and GNSS (ICL-GNSS), 7--9 June 2022, Tampere, Finland\\}
\newcommand{\MYCopyright}{978-1-6654-0575-1/22/\$31.00~\copyright~2022~IEEE}

\makeatletter
\def\ps@headings{%
	\def\@oddhead{}
	\def\@evenhead{}
	\def\@oddfoot{}%
	\def\@evenfoot{}}
\def\ps@IEEEtitlepagestyle{%
	\def\@oddhead{\hfill\MYheader\hfill}
	\def\@evenhead{\hfill\MYheader\hfill}
	\def\@oddfoot{\MYCopyright\hfill}%
	\def\@evenfoot{\MYCopyright\hfill}
    }
\makeatother
    
\renewcommand{\MYheader}{}
\renewcommand{\MYCopyright}{}

\begin{document}

\title{
Towards Accelerated Localization Performance Across Indoor Positioning Datasets \\

\thanks{Corresponding Author: Lucie Klus (\texttt{lucie.klus@tuni.fi})}
\thanks{This work was supported by the European Union’s Horizon 2020 Research and Innovation programme under the Marie Sklodowska Curie grant agreements No. $813278$ (A-WEAR: A network for dynamic wearable applications with privacy constraints, {http://www.a-wear.eu/})  and No.~$101023072$ (ORIENTATE: Low-cost Reliable Indoor Positioning in Smart Factories, {http://orientate.dsi.uminho.pt}).}
}

\author{
    \IEEEauthorblockN{Lucie Klus\IEEEauthorrefmark{1}\textsuperscript{,}\IEEEauthorrefmark{2}, Darwin Quezada-Gaibor\IEEEauthorrefmark{2}\textsuperscript{,}\IEEEauthorrefmark{1}, Joaquín Torres-Sospedra\IEEEauthorrefmark{3},\\ Elena Simona Lohan\IEEEauthorrefmark{1}, Carlos Granell\IEEEauthorrefmark{2} and
    Jari Nurmi\IEEEauthorrefmark{1}}
\IEEEauthorblockA{\IEEEauthorrefmark{1}\textit{Electrical Engineering Unit}, \textit{Tampere University}, Tampere, Finland}
\IEEEauthorblockA{\IEEEauthorrefmark{2}\textit{Institute of New Imaging Technologies}, \textit{Universitat Jaume I}, Castellón, Spain}
\IEEEauthorblockA{\IEEEauthorrefmark{3}\textit{Algoritmi Research Centre}, \textit{Universidade do Minho}, Guimarães, Portugal}
}


\maketitle

\begin{abstract}
The localization speed and accuracy in the indoor scenario can greatly impact the \acl{qoe} of the user.
While many individual machine learning models can achieve comparable positioning performance, their prediction mechanisms offer different complexity to the system.
In this work, we propose a fingerprinting positioning method for multi-building and multi-floor deployments, composed of a cascade of three models for building classification, floor classification, and 2D localization regression.
We conduct an exhaustive search for the optimally performing one in each step of the cascade while validating on $14$ different openly available datasets.
As a result, we bring forward the best-performing combination of models in terms of overall positioning accuracy and processing speed and evaluate on independent sets of samples. We reduce the mean prediction time by $71$\% while achieving comparable positioning performance across all considered datasets. Moreover, in case of voluminous training dataset, the prediction time is reduced down to $1$\% of the benchmark's.
\end{abstract}


\input{sections/1-introduction}
\input{sections/3-methods}

\input{sections/4-results}
\input{sections/5-conclusion}

\balance
\bibliographystyle{IEEEtran}
\bibliography{00-main}

\end{document}

%% file: acronym.tex
\begin{acronym}[XXXXXXXX]
    \acro{1d}[1D]{1-Dimensional}
    \acro{2d}[2D]{2-Dimensional}
    \acro{2dl}[2D-L]{2D Localization}
    \acro{3d}[3D]{3-Dimensional}
    \acro{5g}[5G]{Fifth Generation Mobile Networks}
    \acro{ai}[AI]{Artificial Intelligence}
    \acro{adb}[AdB]{AdaBoost}
    \acro{ae-elm}[AE-ELM]{Auto-Encoder Extreme Learning Machine}
    \acro{aoa}[AoA]{Angle of Arrival}
    \acro{aod}[AoD]{Angle of Departure}
    \acro{ap}[AP]{Access Point}
    \acro{bh}[BH]{Building Hit}
    \acro{ble}[BLE]{Bluetooth Low Energy}
    \acro{cm}[CM]{Common Model}
    \acro{cnn}[CNN]{Convolutional Neural Network}
    \acro{cr}[CR]{Compression Ratio}
    \acro{dbm}[dBm]{decibel-milliwatts}
    \acro{dct}[DCT]{Discrete Cosine Transform}
    \acro{dl}[DL]{Deep Learning}
    \acro{dt}[DT]{Decision Tree}
    \acro{ecdf}[ECDF]{Empirical Cumulative Distribution Function}
    \acro{eeg}[EEG]{electroencephalogram}
    \acro{elm}[ELM]{Extreme Learning Machine}
    \acro{em}[EM]{Encoder Model}
    \acro{fh}[FH]{Floor Hit}
    \acro{fm}[FM]{Full Model}
    \acro{fp}[FP]{fingerprinting}
    \acro{hd}[HDim]{HIDDEN\_DIM}
    \acro{lbs}[LBS]{Location-Based Service}
    \acro{lr}[LRs]{LEARNING_RATES}
    \acro{gda}[GDA]{Generalized Discriminant Analysis}
    \acro{gnss}[GNSS]{Global Navigation Satellite System}
    \acro{ica}[ICA]{Independent Component Analysis}
    \acro{ieee}[IEEE]{Institute of Electrical and Electronics Engineers}
    \acro{iot}[IoT]{Internet of Things}
    \acro{ips}[IPS]{Indoor Positioning System}
    \acro{knn}[$k$-NN]{$k$-Nearest Neighbors}
    \acro{lan}[LAN]{Local Area Network}
    \acro{lbs}[LBS]{Location Based Service}
    \acro{lda}[LDA]{Linear Discriminant Analysis}
    \acro{los}[LoS]{Line of Sight}
    \acro{lr}[LR]{Learning Rate }
    \acro{lsvm}[LSVM]{Linear Support Vector Machine}
    \acro{mac}[MAC]{Media Access Control}
    \acro{mae}[MAE]{Mean Absolute Error}
    \acro{ml}[ML]{Machine Learning}
    \acro{nb}[NB]{Naive Bayes}
    \acro{nlos}[NLoS]{Non-Line-of-Sight}
    \acro{nn}[NN]{Neural Network}
    \acro{pca}[PCA]{Principal Component Analysis}
    \acro{pl}[PL]{Path-Loss}
    \acro{pt}[PT]{Prediction Time}
    \acro{qda}[QDA]{Quadratic Discriminant Analysis}
    \acro{qoe}[QoE]{Quality of Experience}
    \acro{qos}[QoS]{Quality of Service}
    \acro{rf}[RF]{Random Forest}
    \acro{rfid}[RFID]{Radio Frequency Identification Device}
    \acro{rp}[RP]{Reference Point}
    \acro{rs}[RS]{Recommender Systems}
    \acro{rsrp}[RSRP]{Reference Signal Received Power}
    \acro{rss}[RSS]{Received Signal Strength}
    \acro{rssi}[RSSI]{Received Signal Strength Indicator}
    \acro{sota}[SOTA]{State-of-the-art}
    \acro{svm}[SMV]{Support Vector Machine}
    \acro{svd}[SVD]{Singular Value Decomposition}
    \acro{tl}[TL]{Transfer Learning}
    \acro{tlcnn}[TLCNN]{Transfer Learning Convolutional Neural Network}
    \acro{toa}[ToA]{Time of Arrival}
    \acro{tdoa}[TDoA]{Time Difference of Arrival}
    \acro{tsne}[t-SNE]{T-distributed Stochastic Neighbor Embedding}
    \acro{ue}[UE]{User Equipment}
    \acro{uwb}[UWB]{Ultra Wide-Band}
    \acro{wifi}[\mbox{Wi-Fi}]{IEEE 802.11 Wireless LAN}
    \acro{wknn}[W$k$-NN]{Weighted $k$-Nearest Neighbors}
    \acro{wlan}[WLAN]{Wireless LAN}
    \acro{wsn}[WSN]{Wireless Sensors Networks}
  \end{acronym}

%% file: sections/1-introduction.tex
\section{Introduction}
\label{sec:1introduction}
The exponential growth of wearable and \ac{iot} devices requiring positioning services demands a quick response from the \ac{ips} to provide \ac{qoe} to the end-users. The requirement is more impactful in real-time solutions where a fast response is a must~\cite{ometov2021survey}. That is why multiple algorithms and techniques were developed over the years in order to fulfil the requirements of these cutting-edge devices. For instance, data compression techniques have been applied onto indoor positioning data to reduce the processing and response times as well as to improve the positioning accuracy~\cite{klus2020rss}.


Due to the complex signal propagation characteristics in indoor scenarios, fingerprinting methods are commonly utilized. In order to operate, they require a pre-measured dataset of fingerprints, referred to as a radio map, to adapt the matching mechanism to the environment. In large scenarios such as universities, the radio maps contain hundreds of thousands of fingerprints~\cite{MendozaSilva2018longterm, ipin2014ujiindoorloc,tauDBs2021
}, which may be time-consuming to process during the online phase.
To reduce the processing time during the online phase, the authors in \cite{9115419} provided three new methods to enhance the coarse and fine-grained search in $k$-means clustering to be used in the offline and online phase of \ac{wifi} fingerprinting. As a result, the authors reduced the computational cost of estimating the user position, while slightly improving the position estimation.



One of the student teams of the “Data Analytics and \ac{ml}” program's challenge~\cite{rojo2019machine} considered the possibility of cascading the models and achieved the best results among other teams, outperforming several benchmarks. Their performance proves the cascade's efficiency in terms of positioning accuracy, yet does not consider computational effort, nor its generalizaiton capabilities.

The authors of~\cite{s21103418} propose a clustering method based on the strongest \ac{ap} match before performing localization by \ac{knn}. Their solution outperforms benchmark solutions in terms of both \ac{pt} and positioning accuracy.
Similarly, \cite{maneerat2019roc} used classification algorithms to reduce the complexity in the online phase of \ac{wifi} fingerprinting. They included environmental sensors (humidity and temperature sensors) to provide extra information to the fingerprinting techniques, allowing to reduce the location complexity. In the framework of \cite{maneerat2019roc}, two classification techniques were used to first determine the floor where the target node is located, and the second to filter the location of the fingerprint in the radio map. As a result, the authors improved positioning accuracy.

In this work, we develop a novel approach for a multi-building and multi-floor indoor positioning, which takes advantage of the heterogeneity of the labels within the datasets. The proposed cascade of optimally chosen models for building classification, floor classification and 2D localization regression is able to positively impact the floor-hit, as well as highly improve the prediction time of the considered datasets.

The main contributions of this paper are as follows:
\begin{itemize}
    \item We propose an indoor positioning strategy based on the cascade of three \ac{ml} models, designed to iteratively reduce the search-space of the consecutive models and consequently strongly improving the \ac{pt} without harming the positioning capabilities.
    \item We introduce and evaluate $9$ distinct classification models and $7$ different regression model architectures, both underutilized and ever-present ones across the literature, to find a good general well-performing sequence of models on $14$ open-source indoor positioning datasets.
    \item We perform the testing of the proposed cascade model on an independent subset of data, showing a significant increase in efficiency, in terms of \ac{pt}.
\end{itemize}

The rest of the paper is structured as follows. Section~\ref{sec:3methods} introduces the utilized methods and the proposed system model, Section~\ref{sec:4results} presents the datasets, and the numerical results of both the evaluation and testing phases, while Section~\ref{sec:5conclusions} concludes this work.

%% file: sections/3-methods.tex
\section{Materials and Methods}
\label{sec:3methods}


In this section we present the technical parameters of this work, including the utilized \ac{ml} models, and the principles of the fingerprinting method. Also, we describe the proposed system model and its work-flow.

\subsection{Utilized Methods}

In recent decades, the availability of numerous \ac{ml} models became almost abundant, and their utilization omnipresent across scientific branches, as nowadays, almost each scientific paper mentions the advancements in \ac{dl} and \ac{ai}, considering only the most recent models. Nevertheless, the family of \ac{ml} models offers countless different solutions as well, which are hardly ever utilized.

In this work, we utilize $9$ \ac{ml} methods, involving commonly used classifiers and regressors, as well as the under-utilized approaches, such as \ac{lsvm} or \ac{dt}. 

We consider the following classifiers in the evaluation:

\textbf{$\textbf{\textit{k}}$-Nearest Neighbors ($\textbf{\textit{k}}$-NN) } with $k=1$, $3$ and $11$ (further referred to as e.g. $3$NN for \ac{knn} with $k=3$) as the representative of the most commonly used matching algorithm in non-parametric \ac{ips}. We utilize Manhattan distance as the cost function.
The representatives of number of $k$ were chosen based on the comprehensive sweep of values and parameters performed in~\cite{torres2020comprehensive}, also utilizing some of the datasets used in this work. Moreover, the lower values of $k$ perform well in sparsely covered areas, while higher $k$ (e.g. $11$) improves generalization capabilities of the model.

\textbf{Weighted $\textbf{\textit{k}}$-Nearest Neighbors (W$\textbf{\textit{k}}$-NN)} with $k=3$ and $11$ (further referred to as e.g. W$3$NN for Weighted \ac{knn} with $k=3$) is the alteration of \ac{knn}, which after finding the $k$ neighbors considers their importance based on the inverse of their distance to the reference sample. 
The values of $k$ were kept the same as in \ac{knn}, despite \acs{wknn}'s inherent capability of filtering less relevant samples. 

\textbf{\acf{lsvm}} is a \ac{ml} approach capable of efficiently finding boundaries between classes in high-dimensional space. We utilize the model with the linear kernel, as it achieved significantly better performance in terms of both accuracy and the training time than the commonly used Radial Basis Function. The model parameters are 'l2' penalty and the regularization parameter $C$ = $1$. 
    
\textbf{\acf{dt}} creates a tree-like structure of decision boundaries using the most relevant features to perform classification. We apply Gini impurity, $50$  maximum depth of the tree and greedy (best) splitting strategy, since the randomization of trees is performed by the next model. 
    
\textbf{\acf{rf}}, as the name suggests, builds multiple randomized decision trees and performs classification as the ensemble of the individual results. We built the classifier using $10$ trees with $50$ maximum depth, Gini impurity, and the number of features considered in each split limited to a square root of total features. Further increasing the parameters did not provide improved results. 
    
\textbf{\acf{nn}}; The \ac{nn} considered in this work involves of a single hidden densely connected layer with $100$ neurons, followed by a classification layer with the number of neurons equal to the number of classes (e.g. $3$ neurons for $3$ possible floors). The training is realized with Adam optimizer for up to $100$ epochs with `l2' regularization of the hidden layer. We denote, that further tuning of the hyperparameters, which is beyond this paper's scope, might provide improved results, as previously discussed in the literature~\cite{klus2022transfer,song2019novel}.
    
\textbf{\acf{adb}} is an ensemble method, which subsequently grows multiple shallow decision trees, where each conclusive tree focuses on the incorrectly classified samples~\cite{hastie2009multi}. We set the number of estimators to $50$, and depth of each tree to $1$. The idea behind this model is considering the majority vote of many weak classifiers.
    
\textbf{\acf{nb}} is a simple classification scheme, which assumes pair-wise independence of features and their likelihood as Gaussian~\cite{zhang2004optimality}.
    
\textbf{\acf{qda}} fits a Gaussian density to each class, instead of finding only a linear boundary, as in e.g. \ac{lsvm}. The classifier provides improved results in case of intersecting classes.

For the regression, we consider most of the \ac{ml} model architectures as used for the classification schemes. We adjust the models as regressors, allowing to predict continuous values (classification models only predict discrete classes). The regressors include: 
\textbf{$\textbf{\textit{k}}$-Nearest Neighbors ($\textbf{\textit{k}}$-NN) } with $k=1$ \& $k=3$ \& $k=11$, 
\textbf{Weighted $\textbf{\textit{k}}$-Nearest Neighbors (W$\textbf{\textit{k}}$-NN)} with $k=3$ \& $k=11$, 
\textbf{\acf{lsvm}}, \textbf{\acf{dt}}, \textbf{\acf{rf}}, \textbf{\acf{nn}}, and \textbf{\acf{adb}}, with the same hyper-parameter settings as their classifier alternatives.

\subsection{Positioning via Fingerprinting}

Nowadays, fingerprinting is one of the most commonly utilized approaches for localization in \ac{ips}. The simple fact, that the method does not require any specific knowledge about the deployment is one its greatest strengths. Fingerprinting positioning requires only the pre-recorded database of fingerprints from the deployment, and a matching algorithm in order to operate. The approach is most commonly used in deployments with strong \ac{nlos} and signal scattering characteristics, such as indoors~\cite{radar} or dense urban  areas~\cite{lora03,anagnostopoulos2019reproducible,Purohit2020}.

The traditional, and most commonly used matching algorithm is \ac{knn} and its variants, whose tuneable parameters, such as number of $k$ or distance metric, offer great fine-tuning options, resulting in unmatched positioning performance~\cite{torres2020comprehensive}. Recently, many works, such as~\cite{klus2022transfer, song2019novel} substitute \ac{knn} with different types of \acfp{nn}, which ensure faster matching capabilities, especially on large datasets, but usually do not achieve the same positioning accuracy. 

In this work, we aim to evaluate multiple \ac{ml} models, not restricted to \ac{knn} and \ac{nn} only, to build a cascade of models capable of significantly improving the \ac{pt}, as well as keeping or improving the positioning accuracy, compared to using only a single model.

\subsection{Proposed System Model}

In this work, we develop a cascade positioning algorithm, which exploits the abundance of labels within the indoor positioning datasets. The considered cascade of models first finds the building estimate, then estimates the floor, and the last model finds the exact user coordinates, all by considering only the relevant samples. While certain works employ the strategy of creating a model for predicting building, floor, and \ac{2d} position as the separate tasks~\cite{marques2012combining,song2019novel}, we fully utilize the knowledge obtained in each step.

\begin{figure}
    \centering
    \includegraphics[width=1\columnwidth]{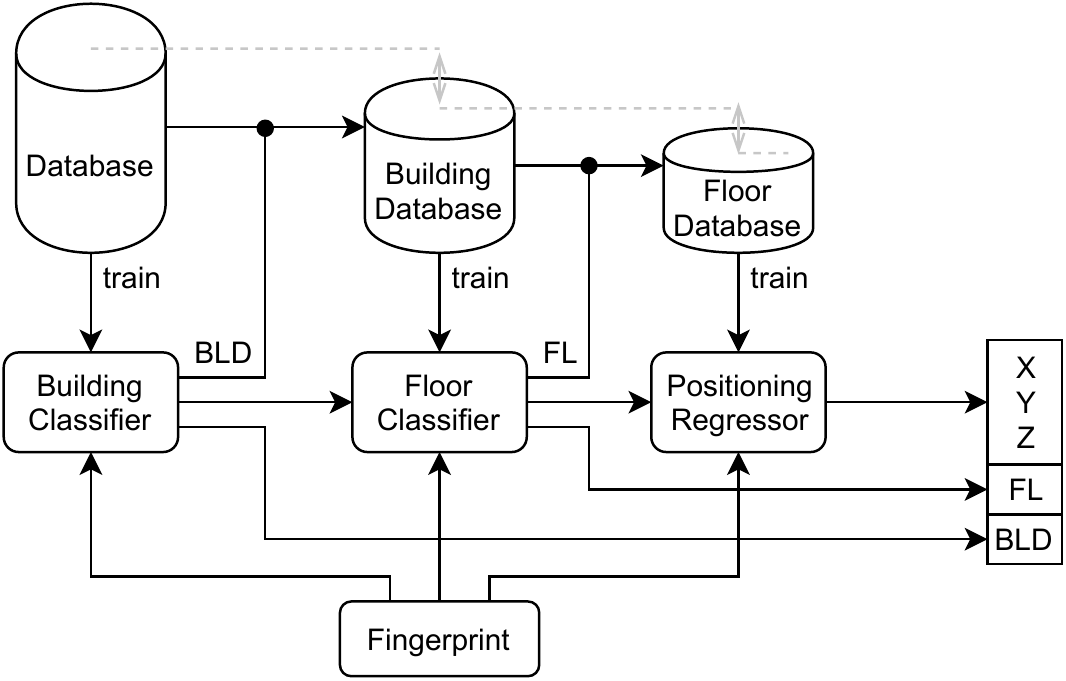}
    \caption{General system model work-flow. Each subsequent model considers only the training samples relevant to the deployed area.}
    \label{fig:algor}
\end{figure}

The visualization of the proposed cascade algorithm is shown in Fig.~\ref{fig:algor}. 
Each dataset first trains a \ac{bh} classifier using all available samples within the deployment, and the building label as the target.
For each building in the deployment, a separate \ac{fh} classifier is created and trained only using the samples from the adequate building, while predicting the floor label.
Finally, for each floor within each building, a separate \ac{2dl} regression model is trained using only the training samples from the relevant floor in the relevant building to accurately localize the user.
After the models are trained in off-line fashion, the incoming fingerprint is evaluated on-line as follows.

In the first stage, the \ac{bh} model estimates the fingerprint's building label, based on which the corresponding trained \ac{fh} classifier is selected. 
In the second stage, the selected \ac{fh} model predicts the fingerprint's floor label and enables the selection of the \ac{2dl} model for the adequate floor.
The third stage of the algorithm predicts the fingerprint's coordinates using the selected model and the ensemble of the building label, floor label and the user coordinates is complete.

The proposed cascade of models offers a stable separation of samples based on building and floor labels into pre-defined number of classes. While \ac{ap}-wise clustering methods~\cite{marques2012combining, maneerat2019roc,s21103418} separate the space into clusters based on samples' features, the proposed cascade splits the data based on their labels and thus ensures more balanced distribution of samples per classifier. As a result, each considered individual prediction model (e.g. per floor) has a sufficient number of samples. The balanced separation is crucial when considering trainable models (e.g. \ac{nn} or \ac{rf}), while reducing the number of training samples accelerates \ac{knn} performance.

In order to demonstrate the best combination of the classifiers and the regressor in the cascaded architecture, we later compare the performance of $9$ commonly known models. We propose the combination of methods to gain comparable accuracy of positioning in comparison to using the traditional model only, while vastly improving the time required for localization (prediction). Additionally, we propose a cascade combination of methods, which further decreases the \ac{pt} at the expense of lowering the positioning accuracy.

%% file: sections/4-results.tex
\section{Experiment and Results}
\label{sec:4results}

In this section, we first introduce the $14$ available datasets. These datasets were split into training and testing subsets and used for tuning, and later  evaluating of the proposed method, as described in this section.

For the implementation of this work, we used Python 3.8 with Numpy, Scipy, and Scikit-Learn libraries. For the result processing and visualization, we used MATLAB R2020b. The hardware included Intel(R) Core(TM) i7-8750H processor with $32$~GB RAM.

For the purposes of reproducibility and replicability of the experiments we share the codes on Zenodo~\cite{zenodo}. 

\subsection{Available Datasets}

In order to evaluate the proposed method and its applicability over different scenarios, we used $14$ publicly available fingerprinting datasets. All of them consist of multiple-floor environments, ranging from $2$ to $16$ floors and in total involving $82$ individual floors, while UJI~1\&2 are also an example of multi-building scenarios (each having 3 buildings). All datasets were obtained using \ac{wifi} as a primary technology for \ac{rss} measurements. More details regarding their properties may be found in Table~\ref{tab:datasets}.

\input{tables/tab-datasets}

The datasets were measured and provided by the three universities.
Universitat Jaume I, Spain provided LIB~1\&2 \cite{MendozaSilva2018longterm} and UJI~1\&2 \cite{ipin2014ujiindoorloc}, 
Tampere University, Finland provided SAH~1 and TIE~1 \cite{tauDBs2021}, TUT~1\&2 \cite{razavi2015k,cramariuc2016clustering}, TUT~3\&4 \cite{LohanTorres-SospedraEtAl17_Wi}, TUT~5 \cite{RichterLohanEtAl18_WLAN} and TUT~6\&7 \cite{taudatasets}.
University of Sydney, Australia provided UTS~1~\cite{song2019novel}.

These datasets were already used in a number of other previously published works, e.g., in \cite{klus2020rss}, \cite{klus2022transfer}, \cite{torres2021towards} and~\cite{quezada2020improving}.


\subsection{Choosing the Methods - Validation}

In this subsection, we explain the method for choosing the best-performing combination of a cascade of two classifiers and a regressor for \ac{bh}, \ac{fh}, and \ac{2dl} estimates, respectively. As a result, we aim to obtain the optimally performing model for an arbitrary dataset.

For this purpose, we split the original training database of each dataset into training and validation parts in 80:20 ratio. The part dedicated for testing is omitted in the whole validation phase, as it is used for unbiased evaluation of the final model only.

\noindent\textbf{\acf{bh}} 

In order to find the most suitable method for building classification, we performed a sweep over all classifiers as described in Section~\ref{sec:3methods}, in terms of validation accuracy, training time and time required for prediction. The results of the sweep can be found in Table~\ref{tab:BHval}, where the validation accuracy is represented by the number of misclassified samples. We provide the total number of misclassifications, rather than a fraction, since only a limited number of samples are in the areas where more than 1 building can be considered. The \ac{pt} reflects the effort required for the task.

\input{tables/tab-BHval}

We note that the performance of the \ac{bh} can only be validated on $2$~datasets, as, among considered, only the datasets UJI~1 and UJI~2 are multi-building environments. 

The results in Table~\ref{tab:BHval} show the perfect classification accuracy of $1$NN, \ac{lsvm} and the \ac{nn} models. As the building misclassification leads to severe positioning errors, therefore the building-hit performance is considered as the most impactful metric. Since the following goal is to reduce the positioning effort, $1$NN classifier was eliminated from consideration due to the lengthy \ac{pt}. Of the remaining two models, \ac{nn} predicted the correct building in $16$~ms, while \ac{lsvm} in $107$~ms, marking it as the best-performing \ac{bh} model. As a downside, according to our evaluation, \ac{nn} required $6.28$~s to train the model, while \ac{lsvm} was trained in $1.34$~s and \ac{knn} does not require any model training at all.
 


\noindent\textbf{\acf{fh}} 

The next validation step finds the most suitable \ac{fh} classifier by performing a sweep over all datasets. The multi-building datasets, UJI~1 and UJI~2,  were split depending on the samples' building labels, forming $6$ smaller databases in total (denoted as e.g. UJI~1$_{1}$ for first building of the UJI~1 dataset). This step effectively follows the $100\%$ \ac{bh} split from the previous step, achieved by any of the three best-performing methods (\ac{knn}, \ac{nn} and \ac{lsvm}).

The results of the performed experiment are presented in Table~\ref{tab:FHval}, showing the full results of the validation accuracy sweep and the averaged results for the \ac{pt}s. The highest \ac{fh} is achieved by \ac{nn} model, followed by \ac{lsvm} and $1$NN. The \ac{pt} of the \ac{nn} is second only to the \ac{dt} model, whose accuracy is lower by $4.3$\%. Consequently, \ac{nn} is chosen as the \ac{fh} classification model.

\input{tables/tab-FHval}

\noindent\textbf{\acf{2dl}} 

In order to find the precise 2D location in the given building and on the given floor, the cascade algorithm runs a 2D regression. To determine the over-all best-performing regression method, we evaluated the results of the sweep across all regression models introduced in Section~\ref{sec:3methods}. The averaged results of the sweep may be found in Table~\ref{tab:2Dval}. We provide only the mean model performance across all floors, as the full table of results (82 floor datasets in total) is too extensive and it does not provide additional value to the paper. 

\input{tables/tab-2Dval}

In all $14$ available and analyzed datasets, there is a total of $82$ floors distributed over $18$ buildings. The overall best positioning accuracy was achieved by $1$NN regression, followed by W$3$NN, with the lowest average positioning error of $2.69$ and $2.93$ meters over all datasets, respectively. Additionally, $1$NN achieved the lowest positioning error for $52$ out of $82$ considered floor datasets and W$3$NN in $20$ floors. The suitability of \ac{knn} for utilization in fingerprinting in terms of good positioning accuracy has been already documented in literature \cite{klus2020rss, torres2020comprehensive, cramariuc2016clustering}, but the time required for prediction is usually very long for voluminous datasets, as it is highly dependent on the datasets size. It is usually considered to be the biggest downside of this method.

Nevertheless, considered \ac{knn} regressors also performed the best in terms of \ac{pt} required to finish the positioning. This is in contrast to \ac{knn}'s behaviour in previous stages (\ac{bh} and \ac{fh} validation), where its \ac{pt} was by far the slowest (as expected) of all considered methods. This change in the behaviour was caused by the notable reduction of the dataset size due to the previous steps in the cascade sequence of the proposed method. 
Additionally, the method does not require any previous training, saving additional resources for the service provider.


\subsection{Numerical Results - Testing}

In this section, we provide the results of the chosen cascade model on the independent test datasets to provide an unbiased evaluation.
\ac{nn} is utilized as the best-performing \ac{bh} and \ac{fh} models, while $1$NN and W$3$NN are utilized as two alternatives of the \ac{2dl} model.
The considered benchmark models are the stand-alone $1$NN and W$3$NN models trained on the full training dataset.
The results of the benchmark directly compare the performance of the cascade model to the stand-alone model with the same parameters. 

Table~\ref{tab:results} presents the results of the proposed solutions, as well as that of the corresponding benchmark models, displaying the raw values for the benchmarks and normalized ones for the proposed solution, directly showing the corresponding improvement or deterioration. We present the raw results for the benchmarks in terms of \ac{bh} [\%], \ac{fh} [\%], \ac{2d} positioning error $\epsilon_{2D}$ [m], \ac{3d} positioning error $\epsilon_{3D}$ [m] and the \ac{pt} [s], while we use the normalized results (denoted as $\Tilde{\ac{bh}}$, $\Tilde{\ac{fh}}$, $\Tilde{\epsilon_{2D}}$, $\Tilde{\epsilon_{3D}}$, and $\Tilde{\ac{pt}}$ [-]) to evaluate the proposed model's performance, similarly to~\cite{torres2021towards}. We denote, that the normalized values are representing the ratios between the proposed method's and the benchmark's results. Therefore, the normalized values lower than $1$, represent the improvement for the entities we aim to decrease ($\epsilon_{2D}$, $\epsilon_{3D}$ and \ac{pt}), while the entities we aim to increase (\ac{bh} and \ac{fh}) are improved in cases, where the normalized value is higher than $1$.
 
The best performing cascade in terms of validation accuracy was the combination of \textbf{\ac{nn}~$\rightarrow$~\ac{nn}}~$\rightarrow$~$\textbf{1NN}$. The overall results show a slight improvement (by $4 \%$) in \ac{fh} and a slight increase of the positioning error ($6 \%$) by the proposed solution while using the same underlying model ($1$NN) as the benchmark. The most relevant improvement is observed in the comparison of the \ac{pt}, which is improved by more than $70$\% compared to the benchmark, on average. Table~\ref{tab:results} also shows, that when the more voluminous datasets are considered, the improvement in \ac{pt} is even more substantial (e.g. the \ac{pt} of the test samples in UJI~1 is reduced from $214$~s to ${2.87}$~s by applying the cascade).

We also provide the averaged normalized results of the second best cascade combination in terms of validation accuracy, namely \ac{nn}~$\rightarrow$~\ac{nn}~$\rightarrow$~W$3$NN. The results are normalized towards W$3$NN benchmark in order to provide unbiased comparison to the plain use of the single model. Although, in comparison with the previous combination, the overall \ac{fh} has been further improved (by an additional $1\%$ due to more favorable random initialization) and the \ac{pt} has decreased $6$\% more, the average positioning accuracy has dropped by $2$\%, adding up to the $8$\% drop in \ac{3d} positioning error in comparison to the use of $1$NN regression. Although the performance of the two is comparable, the trade-off is unfavorable in general.

Additionally, we offer a cascade model combination of \ac{dt}~$\rightarrow$~\ac{dt}~$\rightarrow$~W$3$NN, which represents the combination of the fastest methods in the validation. This cascade provides a drastic decrease in \ac{pt}, as it requires only $16$\% of time in comparison to the W$3$NN benchmark, on average. Specifically, the \ac{pt} of the aforementioned test set in UJI~1 is only $2.38$~s. Nevertheless, this cascade increases the positioning error by $25$\% compared to the benchmark, which may not be an acceptable trade-off for many applications. 

\input{tables/tab-results}

\begin{figure}
    \centering
    \includegraphics[width=0.95\columnwidth]{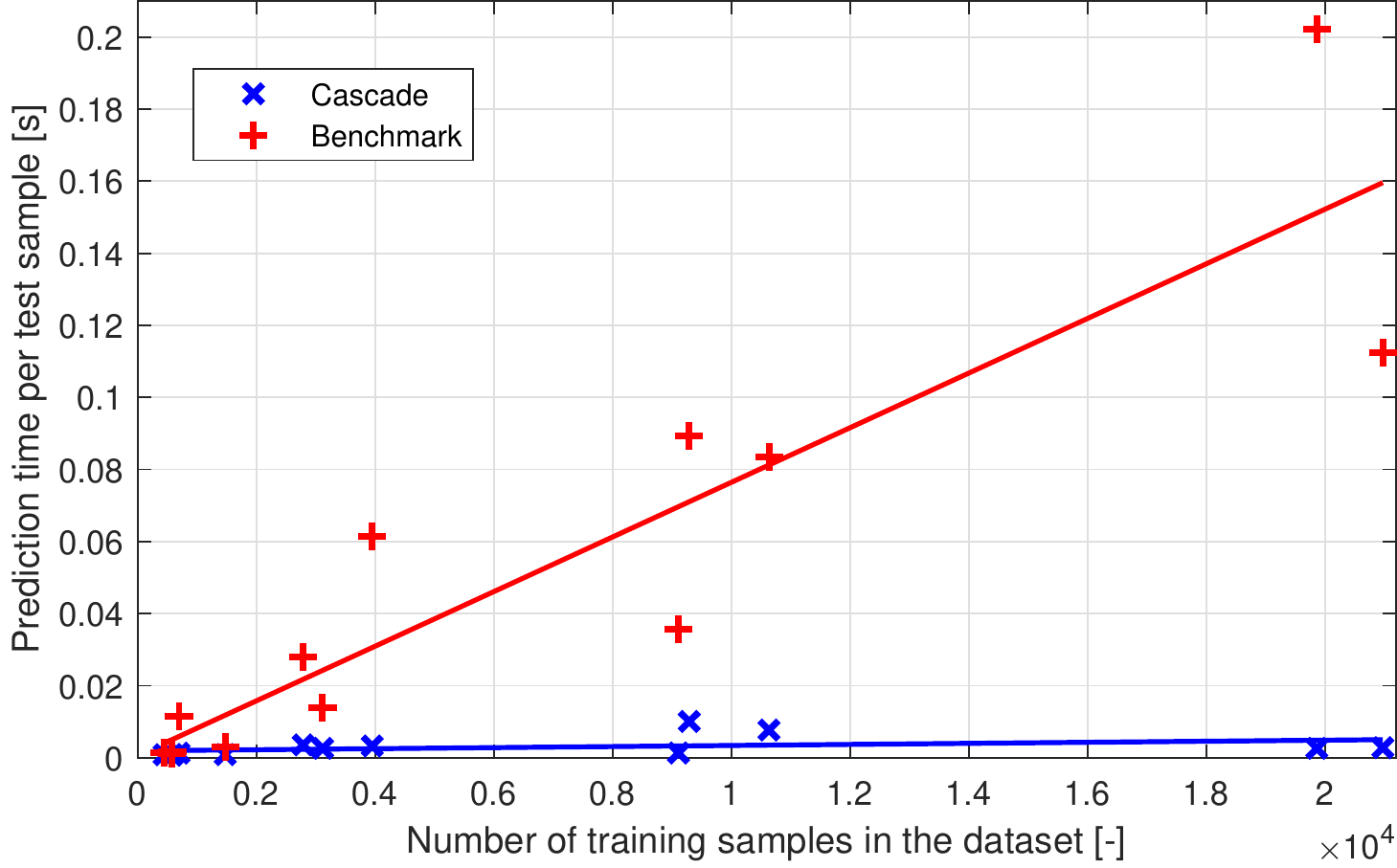}
    \caption{Comparison of a test sample prediction speed between the \ac{nn}~$\rightarrow$~\ac{nn}~$\rightarrow$~$1$NN model and the $1$NN benchmark.}
    \label{fig:times}
\end{figure}

The dependency of the \ac{pt} on the number of samples in the training dataset is visualized in Fig.~\ref{fig:times}, along with their linear expectation. The prediction time of the proposed cascade model (blue in Fig.~\ref{fig:times}) splits each database into smaller parts, thus effectively reduces the \ac{knn}'s search space and consequently the \ac{pt} of the solution. While the results of the benchmark show the steady increase in the prediction time with the increasing volume of the training dataset, the number of samples itself is not the only parameter affecting the \ac{knn}'s complexity, as e.g. the number of \ac{ap}s and the chosen distance metric impact the complexity of the calculation as well. 

\subsection{Discussion}

Section~\ref{sec:4results} first presents a sweep over numerous \ac{ml} methods, then based on the performance, we choose the best performing ones. The procedure is logical, and the final models are the representatives of the most popular models from the literature, namely \ac{nn} or \ac{knn}. The logical conclusion is that their ever-present utilization is well-justified. The \ac{nn}s offer the unmatched generalization properties, while \ac{knn} and its alternatives offer straightforward matching approach. Nevertheless, the remaining models' performance was closely behind the chosen ones (\ac{nn} and \ac{knn}), outperforming them in several evaluation metrics, such as \ac{dt} having the fastest \ac{pt}. The results of this work prove the effectiveness of the mainstream \ac{ml} methods, while highlighting the vast options a data scientist has in terms of model variety. Additionally, we show that by utilizing the models in cascade to iteratively filter the samples, the \ac{2dl} $1$NN model's \ac{pt} can be drastically reduced.

%% file: tables/tab-datasets.tex
\begin{table}[!t]
    \tabcolsep 3pt
    
    \caption{Overview of Datasets}
    \label{tab:datasets}
    \centering
    \begin{adjustbox}{width=\columnwidth,center}
    \begin{tabular}{l cc cc cc}
       \toprule
        Dataset 
        & Samples
        & Training samples
        & Testing samples
        & APs
        & Buildings
        & Floors  \\
        \midrule
LIB~1	&	3696	&	576	    &	3120	&	174	&	1	&	2	\\
LIB~2	&	3696	&	576	    &	3120	&	197	&	1	&	2	\\
SAH~1	&	9447	&	9291	&	156	    &	775	&	1	&	3	\\
TIE~1	&	10683	&	10633	&	50	    &	613	&	1	&	6	\\
TUT~1	&	1966	&	1476	&	490	    &	309	&	1	&	4	\\
TUT~2	&	760	    &	584	    &	176	    &	354	&	1	&	3	\\
TUT~3	&	4648	&	697	    &	3951	&	992	&	1	&	5	\\
TUT~4	&	4648	&	3951	&	697	    &	992	&	1	&	5	\\
TUT~5	&	1428	&	446	    &	982	    &	489	&	1	&	3	\\
TUT~6	&	10385	&	3116	&	7269	&	652	&	1	&	4	\\
TUT~7	&	9291	&	2787	&	6504	&	801	&	1	&	3	\\
UJI~1	&	20972	&	19861	&	1111	&	520	&	3	&	13	\\
UJI~2	&	26151	&	20972	&	5179	&	520	&	3	&	13	\\
UTS~1	&	9496	&	9108	&	388	    &	589	&	1	&	16	\\

         \bottomrule
    \end{tabular}
 \end{adjustbox}  
\end{table}

%% file: tables/tab-BHval.tex
\begin{table}[!t]
    \tabcolsep 3pt
    
    \caption{\acl{bh} Classification Results for Validation}
    \label{tab:BHval}
    \centering
    \begin{adjustbox}{width=\columnwidth,center}
    \begin{tabular}{l cc cc cc cccccc}
       \toprule
        & $1$NN
        & $3$NN
        & W$3$NN
        & $11$NN
        & W$11$NN
        & \ac{lsvm}
        & \ac{dt}
        & \ac{rf}
        & \ac{nn}
        & \ac{adb}
        & \ac{nb}
        & \ac{qda}  \\
        \midrule
        \multicolumn{13}{c}{\textbf{Validation Misclassification [samples]}} \\
UJI~1&  \textbf{0}	&	\textbf{0}	&	\textbf{0}	&	1	&	1	&	\textbf{0}	&	2	&	1	&	\textbf{0}	&	190	&	3	&	73 \\
UJI~2&  \textbf{0}	&	2	&	2	&	3	&	3	&	\textbf{0}	&	5	&	5	&	\textbf{0}	&	45	&	7	&	24 \\
         \midrule
Avg  &  \textbf{\textcolor{Green}{0}}	&	1	&	1	&	2	&	2	&	\textbf{\textcolor{Green}{0}}	&	3	&	3	&	\textbf{\textcolor{Green}{0}}	&	117	&	5	&	48 \\
         \bottomrule \\
        \multicolumn{13}{c}{\textbf{Validation Prediction Time [s]}} \\ 
UJI~1&  27.31	&	27.39	&	25.81	&	26.00	&	26.23	&	0.103	&	\textbf{0.007}	&	0.012	&	0.014	&	0.238	&	0.051	&	0.104 \\
UJI~2&  27.78	&	28.60	&	30.87	&	30.04	&	30.72	&	0.110	&	\textbf{0.008}	&	0.013	&	0.017	&	0.270	&	0.059	&	0.110 \\
         \midrule
Avg &   27.54	&	27.99	&	28.34	&	28.02	&	28.48	&	0.107	&	\textbf{\textcolor{Green}{0.008}}	&	0.013	&	\textbf{\textcolor{Green}{0.016}}	&	0.254	&	0.055	&	0.107 \\
         \bottomrule
    \end{tabular}
 \end{adjustbox}  
\end{table}

%% file: tables/tab-FHval.tex
\begin{table}[!t]
    \tabcolsep 1.20pt
    
    \caption{\acl{fh} Classification Results for Validation}
    \label{tab:FHval}
    \centering
    \begin{adjustbox}{width=\columnwidth,center}
    \begin{tabular}{l cc cc cc cccccc}
       \toprule
        & $1NN$
        & $3$NN
        & W$3$NN
        & $11$NN
        & W$11$NN
        & \ac{lsvm}
        & \ac{dt}
        & \ac{rf}
        & \ac{nn}
        & \ac{adb}
        & \ac{nb}
        & \ac{qda}  \\
        \midrule
        \multicolumn{13}{c}{\textbf{Validation Accuracy [\%]}} \\
LIB~1&	\textbf{100}	    &	\textbf{100}	    &	\textbf{100}	    &	\textbf{100}	    &	\textbf{100}	    &	\textbf{100}	    &	98.3	&	\textbf{100}	    &	\textbf{100}	    &	\textbf{100} 	&	96.6	&	91.4	\\
LIB~2&	\textbf{100}	    &	\textbf{100}	    &	\textbf{100}	    &	\textbf{100}	    &	\textbf{100}	    &	\textbf{100}	    &	99.1	&	\textbf{100}	    &	\textbf{100}	    &	\textbf{100} 	&	97.4	&	97.4	\\
SAH~1&	\textbf{100}	    &	\textbf{100}	    &	\textbf{100}	    &	99.8	&	\textbf{100}	    &	\textbf{100}	    &	99.7	&	99.6	&	\textbf{100} 	&	88.6	&	96.0	&	96.3	\\
TIE~1&	\textbf{100}	    &	99.9	&	99.9	&	99.6	&	99.7	&	\textbf{100}	    &	99.5	&	99.4	&	99.9	&	67.7	&	87.4	&	87.4	\\
TUT~1&	98.0	&	98.3	&	98.3	&	97.0	&	97.3	&	99.0	&	93.6	&	96.6	&	\textbf{99.3}	&	66.2	&	56.1	&	54.1	\\
TUT~2&	\textbf{100}	    &	99.1	&	99.1	&	94.0	&	97.4	&	98.3	&	96.6	&	97.4	&	99.1	&	89.7	&	74.4	&	71.8	\\
TUT~3&	89.3	&	82.1	&	82.9	&	72.1	&	75.7	&	93.6	&	73.6	&	91.4	&	\textbf{95.0}	&	55.7	&	60.7	&	42.9	\\
TUT~4&	95.7	&	96.1	&	95.8	&	94.2	&	94.4	&	96.3	&	87.5	&	94.3	&	\textbf{96.6}	&	66.9	&	61.3	&	69.9	\\
TUT~5&	\textbf{100}	    &	\textbf{100}	    &	\textbf{100}	    &	90.0	&	95.6	&	\textbf{100}	    &	94.4	&	92.2	&	\textbf{100}	    &	62.2	&	91.1	&	75.6	\\
TUT~6&	99.8	&	\textbf{100}	    &	\textbf{100}	    &	99.8	&	99.8	&	\textbf{100}	    &	97.3	&	99.2	&	\textbf{100} 	&	71.2	&	84.6	&	84.1	\\
TUT~7&	\textbf{98.9}	&	97.8	&	98.6	&	97.7	&	97.8	&	98.4	&	96.1	&	96.8	&	98.6	&	91.2	&	92.7	&	85.5	\\
UJI~1$_{1}$&	\textbf{99.8}	&	99.7	&	99.7	&	99.4	&	99.5	&	99.5	&	97.2	&	99.1	&	\textbf{99.8}	&	63.4	&	49.8	&	68.1	\\
UJI~1$_{2}$&	\textbf{99.9}	&	99.8	&	99.8	&	99.2	&	99.7	&	99.6	&	98.4	&	99.6	&	99.7	&	68.1	&	63.7	&	64.6	\\
UJI~1$_{3}$&	\textbf{99.8}	&	99.6	&	\textbf{99.8}	&	99.6	&	99.7	&	99.4	&	97.9	&	99.3	&	99.6	&	56.4	&	47.0	&	42.0	\\
UJI~2$_{1}$&	98.7	&	98.8	&	99.0	&	98.7	&	98.8	&	\textbf{99.1}	&	96.0	&	98.3	&	\textbf{99.1}	&	58.7	&	42.8	&	48.6	\\
UJI~2$_{2}$&	99.1	&	98.7	&	98.8	&	98.2	&	98.6	&	\textbf{99.2}	&	96.6	&	98.4	&	\textbf{99.2}	&	68.7	&	65.7	&	64.1	\\
UJI~2$_{3}$&	99.6	&	99.3	&	99.4	&	99.5	&	99.6	&	99.5	&	97.4	&	98.8	&	\textbf{99.7}	&	60.8	&	51.5	&	48.4	\\
UTS~1&	\textbf{99.9}	&	99.8	&	\textbf{99.9}	&	99.3	&	\textbf{99.9}	&	\textbf{99.9}	&	99.5	&	99.8	&	\textbf{99.9}	&	23.1	&	92.3	&	98.8	\\

         \midrule
Avg  &  98.8	&	98.3	&	98.4	&	96.6	&	97.4	&	99.0	&	95.5	&	97.8	&	\textbf{\textcolor{Green}{99.2}}	&	69.9	&	72.8	&	71.7	\\ 
         \bottomrule \\
        \multicolumn{13}{c}{\textbf{Validation Prediction Time [s]}} \\ 
Avg &   2.687	&	2.675	&	2.657	&	2.724	&	2.622	&	0.052	&	\textbf{\textcolor{Green}{0.002}}	&	0.005	&	\textbf{\textcolor{Green}{0.003}}	&	0.084	&	0.026	&	0.056 \\
         \bottomrule
    \end{tabular}
 \end{adjustbox}  
\end{table}

%% file: tables/tab-2Dval.tex
\begin{table}[!t]
    \tabcolsep 3pt
    
    \caption{2D Regression Results for Validation}
    \label{tab:2Dval}
    \centering
    \begin{adjustbox}{width=\columnwidth,center}
    \begin{tabular}{l cc cc cc cccc}
       \toprule
        & $1NN$
        & $3$NN
        & W$3$NN
        & $11$NN
        & W$11$NN
        & \ac{lsvm}
        & \ac{dt}
        & \ac{rf}
        & \ac{nn}
        & \ac{adb} \\
        \midrule
        \multicolumn{11}{c}{\textbf{Validation Accuracy [m]}} \\
Avg	&	\textbf{\textcolor{Green}{2.69}}	&	3.27	&	\textbf{2.93}	&	5.64	&	4.18	&	8.90	&	4.03	&	4.90	&	5.04	&	8.06	\\
         \midrule \\
        \multicolumn{11}{c}{\textbf{Validation Prediction Time [ms]}} \\ 
Avg	&	\textbf{\textcolor{Green}{0.08}}	&	\textbf{0.08}	&	\textbf{\textcolor{Green}{0.07}}	&	\textbf{0.07}	&	\textbf{0.08}	&	77.4	&	5.31	&	7.08	&	180.2	&	111.2	\\
         \midrule \\
        \multicolumn{11}{c}{\textbf{Number of floor datasets with the lowest \ac{2d} pos. error [-]}} \\ 
&	\textbf{\textcolor{Green}{52}}	&	1	&	\textbf{20}	&	0	&	2	&	0	&	5	&	0	&	1	&	1	\\
        
         \bottomrule
    \end{tabular}
 \end{adjustbox}  
\end{table}

%% file: tables/tab-results.tex
\begin{table}[!t]
    \tabcolsep 3pt
    
    \caption{Numerical results of the benchmark and the proposed cascade of models}
    \label{tab:results}
    \centering
    \begin{tabular}{l cc SS cc Sccc}
       \toprule
        & \ac{bh}
        & \ac{fh}
        & $\epsilon_{2D}$
        & $\epsilon_{3D}$
        & \ac{pt} 
        & $\Tilde{\ac{bh}}$
        & $\Tilde{\ac{fh}}$
        & $\Tilde{\epsilon_{2D}}$
        & $\Tilde{\epsilon_{3D}}$
        & $\Tilde{\ac{pt}}$  \\
        
        & [\%] & [\%] & [m] & [m] & [s] & [-] & \text{[-]} & [-] & [-] & [-]  \\
        \midrule
        &\multicolumn{5}{c}{Baseline}&\multicolumn{5}{c}{\textbf{\ac{nn}~$\rightarrow$~\ac{nn}}~$\rightarrow$~$\textbf{1NN}$} \\
LIB~1&			&	99.84	&	3.03	&	3.04	&	2.91	&		&	1	    &	1.11	&	1.10	&	0.82	\\
LIB~2&			&	97.72	&	4.13	&	4.20	&	3.25	&		&	1.02	&	0.93	&	0.91	&	0.83	\\
SAH~1&			&	46.79	&	8.26	&	9.05	&	13.6	&		&	1.26	&	0.97	&	0.95	&	0.12	\\
TIE~1&			&	60.00	&	6.20	&	7.16	&	3.94	&		&	1	    &	0.92	&	0.92	&	0.10	\\
TUT~1&			&	90.00	&	9.50	&	9.60	&	1.24	&		&	1	    &	1.03	&	1.03	&	0.35	\\
TUT~2&			&	72.73	&	12.71	&	12.89	&	0.22	&		&	1.10	&	1.23	&	1.23	&	0.65	\\
TUT~3&			&	91.62	&	9.49	&	9.59	&	42.8	&		&	1.03	&	1.05	&	1.04	&	0.12	\\
TUT~4&			&	95.27	&	6.34	&	6.41	&	42.7	&		&	1.01	&	1.07	&	1.07	&	0.05	\\
TUT~5&			&	88.39	&	6.84	&	6.92	&	1.41	&		&	1.08	&	1.18	&	1.17	&	0.60	\\
TUT~6&			&	99.99	&	1.96	&	1.96	&	96.7	&		&	1	    &	1.06	&	1.06	&	0.20	\\
TUT~7&			&	99.19	&	2.34	&	2.35	&	178	    &		&	1	    &	1.02	&	1.02	&	0.13	\\
UJI~1&		100	&	87.76	&	10.73	&	10.83	&	214	    &	1	&	1.03	&	0.90	&	1.04	&	0.01	\\
UJI~2&		100	&	85.34	&	7.87	&	8.05	&	572	    &	1	&	1.02	&	1.20	&	1.19	&	0.03	\\
UTS~1&			&	92.78	&	8.38	&	8.76	&	13.1	&		&	1.04	&	1.09	&	1.05	&	0.04	\\
         \midrule 
Avg	&		&		&		&		&		&	1	&	1.04	&	1.06	&	1.06	&	0.29	\\
         \bottomrule \\
        &\multicolumn{5}{c}{}&\multicolumn{5}{c}{\textbf{\ac{nn}~$\rightarrow$~\ac{nn}}~$\rightarrow$~$\textbf{W3NN}$} \\
Avg	&&		&		&		&		& 	1	&	1.05	&	1.09	&	1.08	&	0.23	\\
         \bottomrule \\
        &\multicolumn{5}{c}{}&\multicolumn{5}{c}{\textbf{\ac{dt}~$\rightarrow$~\ac{dt}}~$\rightarrow$~$\textbf{W3NN}$} \\
Avg	&		&		&		&		&		&	1	&	0.96	&	1.25	&	1.25	&	0.16	\\
         \bottomrule
    \end{tabular}
\end{table}

%% file: sections/5-conclusion.tex
\section{Conclusion}
\label{sec:5conclusions}

This work proposes an efficient strategy for accelerating the \ac{rss} positioning in indoor scenarios. The proposed solution utilizes a cascade of \ac{ml} models to perform the positioning task, while drastically saving computational resources and achieving comparable positioning accuracy to the benchmark. The proposed system model utilizes the building and floor labels to iteratively reduce the search space of the consecutive models.

In the evaluation, we performed the comparison of a wide variety of \ac{ml} models for each task, including $9$ distinct classifiers and $7$ regressors, while comparing their performance on $14$ datasets. As the result of the exhaustive evaluation, the work distinguished the cascade of neural network (i.e., the first stage in the cascaded architecture), neural network (second stage) and $1$NN (third stage) for building classification, floor classification, and \ac{2d} localization, respectively, as the combination with the best positioning performance. The second highlighted model was a cascade of two decision trees for the first two stages and weighted $3$NN for the third cascaded stage, as the fastest-performing combination. 

The results on the independent test sets showed that the strategy of sequentially splitting the voluminous datasets enables absolute freedom in selecting the optimal model for each task, while the \ac{2d} localization model still efficiently utilizes the highly accurate \ac{knn}, without the drawback of a lengthy prediction. The proposed best-performing model reduced the \acl{pt} by $71$\% and improved the \acl{fh} by $4$\%, on average, while only slightly deteriorating  the \ac{2d} positioning error, when compared to the stand-alone $1$NN benchmark. The results for the voluminous dataset UJI~1 demonstrated $100$-fold decrease in prediction time. Additionally, the fastest-performing model was able to reduce the \acl{pt} by $84$\% on average, at the cost of $4$\% lower \acl{fh} and $25$\% lower positioning accuracy.